\documentclass[10pt,twocolumn]{IEEEtran}
\usepackage{fancyhdr}
\usepackage{cite}
\usepackage{amsmath,amssymb,amsfonts}
\usepackage{optidef}
\usepackage{algorithmic}
\usepackage{graphicx}
\usepackage{textcomp}
\usepackage{soul}
\DeclareMathOperator*{\minimize}{min}

\DeclareMathOperator*{\logb}{logb}
\DeclareMathOperator*{\llogb}{logb^*}
\DeclareMathOperator*{\diag}{diag}
\newcommand{\gbar}{\text{\st{$\gamma$}}}
\def\BibTeX{{\rm B\kern-.05em{\sc i\kern-.025em b}\kern-.08em
    T\kern-.1667em\lower.7ex\hbox{E}\kern-.125emX}}
\usepackage{etoolbox}
\makeatletter
\@ifundefined{color@begingroup}%
  {\let\color@begingroup\relax
   \let\color@endgroup\relax}{}%
\def\fix@ieeecolor@hbox#1{%
  \hbox{\color@begingroup#1\color@endgroup}}
\patchcmd\@makecaption{\hbox}{\fix@ieeecolor@hbox}{}{\FAILED}
\patchcmd\@makecaption{\hbox}{\fix@ieeecolor@hbox}{}{\FAILED}
\usepackage{xcolor}   
\newcommand{\rev}[1]{{\color{black} #1}} 
\begin{document}
\title{OPTIKS: Optimized Gradient Properties Through Timing in K-Space}
\author{Matthew A. McCready, Xiaozhi Cao, Kawin Setsompop, John M. Pauly, \IEEEmembership{Fellow, IEEE}, and Adam B. Kerr
\thanks{Manuscript recieved March 5, 2025; revised Sept 5, 2025. This work was supported in part by the National Institutes of Health (NIH) under Grant R01 EB009690, and Grant U01 EB029427, as well as funding from the Natural Sciences and Engineering Research Council of Canada (NSERC) PGS-D and GE Healthcare.}
\thanks{M. A. McCready and A. B. Kerr are with the Department of Electrical Engineering and the Stanford Center for Cognitive and Neurobiological Imaging, Stanford University, Stanford, CA 94305 USA (e-mail: mattmc@stanford.edu, akerr@stanford.edu). }
\thanks{X. Cao amd K. Setsompop are with the Department of Radiology, Stanford University, Stanford, CA 94305 USA (e-mail: xiaozhic@stanford.edu, kawins@stanford.edu). }
\thanks{J. M. Pauly, is with the Department of Electrical Engineering, Stanford University, Stanford, CA 94305 USA (e-mail: pauly@stanford.edu).}
\thanks{© 2025 IEEE.  Personal use of this material is permitted.  Permission from IEEE must be obtained for all other uses, in any current or future media, including reprinting/republishing this material for advertising or promotional purposes, creating new collective works, for resale or redistribution to servers or lists, or reuse of any copyrighted component of this work in other works}
\thanks{Citation Information: DOI 10.1109/TMI.2025.3639398}}

\maketitle

\begin{abstract}
A customizable method (OPTIKS) for designing fast trajectory-constrained gradient waveforms with optimized time domain properties was developed. Given a specified multidimensional \textit{k}-space trajectory, the method optimizes traversal speed (and therefore timing) with position along the trajectory. OPTIKS facilitates optimization of objectives dependent on the time domain gradient waveform and the arc-length domain \textit{k}-space speed. OPTIKS is applied to design waveforms which limit peripheral nerve stimulation (PNS), minimize mechanical resonance excitation, and reduce acoustic noise. A variety of trajectory examples are presented including spirals, circular echo-planar-imaging, and rosettes. Design performance is evaluated based on duration, standardized PNS models, field measurements, gradient coil back-EMF measurements, and calibrated acoustic measurements. We show reductions in back-EMF of up to 94\% and field oscillations up to 91.1\%, acoustic noise decreases of up to 9.22 dB, and with efficient use of PNS models speed increases of up to 11.4\%. The design method implementation is made available as an open source Python package through GitHub (https://github.com/mamccready/optiks).
\end{abstract}


\section{Introduction}
\label{sec:introduction}
\IEEEPARstart{H}{igh} performance gradient hardware has seen growing popularity in recent years, in order to facilitate fast high-resolution MRI \cite{hardware_hp, UHF_vibrations}. Demand for fast imaging, and improvements in image reconstruction have also led to a wide range of non-Cartesian \textit{k}-space sampling trajectories \cite{spiral_epi, VD_spiral, rosette, cepi, SPARKLING, BJORK, trajopt_pns}. With increased gradient amplitudes, faster slew-rates, and asymmetric coil designs comes greater risk to patient and system safety.

Rapidly varying magnetic fields can cause patients to experience peripheral nerve stimulation (PNS), a major source of discomfort in rapid imaging \cite{pns_1, pns_2}. Limits on PNS threshold are imposed by an international engineering standard (IEC 60601-2-33) \cite{IEC60601} and apply to every scan of a human with an MRI. The maximum PNS limit ($P_{max}$) often prevents operating the gradient system at its maximum slew-rate ($S_{max}$) and by extension maximum gradient amplitude ($G_{max}$).

Gradient waveforms are often globally de-rated to a slew-rate much less than $S_{max}$ to meet PNS limits thus resulting in longer scan times. Previous work has optimized fast PNS approved spirals by introducing a $P_{max}$-limited region in addition to existing $G_{max}$ and $S_{max}$-limited regions \cite{pns_spiral}. Faster EPI sequences have also met $P_{max}$ by empirically replacing ramp periods with $tanh$ functions \cite{pns_spiral}. Methods have also been presented for optimization of \textit{k}-space trajectory constrained by $P_{max}$, but these cannot follow a pre-specified trajectory and are designed for a user-chosen readout time with no option for time optimization \cite{trajopt_pns}.

Rapid variation of gradient fields is also the source of gradient coil vibration. Playing out a gradient waveform modulates the magnitude and direction of current flow in the coils and therefore the Lorentz forces they experience. This drives vibrations and deformations of the coils in tandem with the gradient waveform. In particular, if a gradient waveform contains power at the mechanical resonance frequencies of the coils it will excite those resonances, producing strong and long lasting vibrations. This can result in permanent damage to the coils and produce oscillatory error fields \cite{hardware_hp, UHF_vibrations}. Furthermore, eddy currents induced on conducting surfaces in the scanner by the gradients will experience these vibrations. Components close to the liquid helium bath, such as the radiation shield, can deposit large amounts of energy into the helium causing it to boil off when vibrated \cite{UHF_vibrations, he_bath}. This loss of helium is expensive to replace and can put the system in danger of quenching posing a major safety risk to both system and patient. Limited methods have been presented for producing compact spectrum spiral waveforms \cite{spirals_bandlimited}, and frequency controlled waveforms with fixed timing and a set zeroth moment \cite{synaptive_patent}. A heuristic method has been developed for reducing spiral waveform power in specified bands based on control of an assumed instantaneous frequency \cite{safespiral}, but does not take into account transient effects and in some cases \textit{increases} mechanical resonance vibrations. Currently there is no method for explicit gradient frequency control while following a specified trajectory.

Gradient waveform driven vibrations are also responsible for loud acoustic noise present in MRI exams. This is the number one source of patient discomfort and had been the focus of many studies to make sequences quieter. Current methods to design quiet gradient waveforms severely band-limit the waveforms \cite{acoustic_soft, acoustic_spiral}. These methods are highly restricted in applicable trajectories, and lack finer frequency control to cancel out loud vibration bands below the set band-limit. Scanner specific acoustic transfer functions (ATFs) have been measured under a linear-time-invariant assumption to predict acoustic noise based on the input gradient waveform \cite{acousticLTI_fMRI, acoustic_spatial}. A method for gradient frequency control could use ATFs to efficiently reduce acoustic noise while following specified readout trajectories.

A time-optimal method for gradient waveforms which follow arbitrary trajectories is developed in \cite{miki_topt}. A framework is presented where gradient waveforms are designed by optimizing speed for following an arbitrary \textit{k}-space trajectory, as a function of distance along the curve (arc-length). This is used to find waveforms which trace out \textit{k}-space trajectories in minimum time adhering to $G_{max}$ and $S_{max}$, an optimal control problem with an explicit solution. While efficient, the method is highly inflexible with regard to inclusion of time-domain waveform properties.

In this work we present a method to design gradient waveforms ($\mathbf{g}(t)$) with optimized properties by modifying timing in k-space (OPTIKS) for any prescribed trajectory. We expand upon the \textit{k}-space speed approach developed in \cite{miki_topt} by replacing the explicit time-optimal solution with an iterative gradient descent optimization. This method allows the optimization of \textit{k}-space speed as a function of trajectory arc-length $v(s)$ that minimizes a loss function $L(v(s),\mathbf{g}(t\{v(s)\}))$, bridging the gap between time-domain waveform optimization objectives and the optimal arc-length domain speed. We then apply this new method to design fast gradient waveforms which adhere to $G_{max}$ and $S_{max}$, while for the first time allowing for PNS, and frequency control without trajectory deviation. We present examples including spiral, rosette, and circular echo-planar-imaging (CEPI) trajectories \cite{spiral_epi, rosette, cepi}. The method is made available as an open source Python package through GitHub (https://github.com/mamccready/optiks).

\section{Methods}
\label{sec:methods}
Given a k-space trajectory in an arbitrary parameterization $\mathbf{C}(p)=(x(p), y(p), z(p))$  it can always be parameterized in terms of its Euclidean arc-length $s$. Calculating $s$ from the derivative $\mathbf{C}'(p)$,
\begin{equation}
    s(p)=\int_{0}^{p}\lVert \mathbf{C}'(q) \rVert_2 dq
\label{arc-length}\end{equation}
and inverting and substituting back into $\mathbf{C}(p)$ gives $\mathbf{C}(s)$. The choice of speed in \textit{k}-space at each arc-length $v(s)$ along the trajectory then determines the timing for passing along the curve,
\begin{equation}
    t(s)=\int_{0}^{s}{d\sigma/v(\sigma)}.
\label{time}\end{equation}
With timings $t(s)$ determined, inverting gives arc-length as a function of time $s(t)$ and substitution into $\mathbf{C}(s)$ gives the time parameterization of the trajectory $\mathbf{C}(t)$. The gradient $\mathbf{g}(t)$ is then directly calculated via time differentiation of $\mathbf{C}(t)$ \eqref{goft}
\begin{equation}
    \mathbf{g}(t)=\frac{1}{\gbar}\frac{d\mathbf{C}(s(t))}{dt}.
\label{goft}\end{equation}
Lustig et al explicitly solve \eqref{arcopt} \cite{miki_topt} for the $v(s)$ which minimizes waveform duration with constraints on $v(s)$ and its derivative with respect to arc-length $v'(s)$. These constraints enforce the limits on gradient amplitude and slew-rate in the arc-length domain and incorporate the length $L$ trajectory geometry through its curvature $\kappa(s)=\lVert \mathbf{C}''(s) \rVert_2$,
\begin{equation}
    \begin{aligned}
        \minimize_{v(s)} \quad & \int_{0}^{L}{ds/v(s)}\\
        \textrm{subject to} \quad & v(s)\le\min{\left\{\gbar G_{max},\sqrt{\gbar S_{max}/\kappa\left(s\right)}\right\}\ }\\
        & |v^\prime(s)|\le\frac{1}{v(s)}\sqrt{\gbar^2{S_{max}}^2-\kappa(s)^2v(s)^4}\\
        & v(0)=0.
    \end{aligned}
\label{arcopt}\end{equation}
The problem we seek to solve here is rooted in the same optimization – optimizing speed for arbitrary trajectories – but with the goal of optimizing some time domain property of the resulting gradient waveform. The problems presented here do not have explicit solutions and will be solved using back-propagation with the iterative gradient descent optimizer “Adam” included in PyTorch.

\subsection{Differentiable Domain Transformations}
Extending the optimization problem to time domain waveform properties requires a differentiable conversion from $v(s)$ to $\mathbf{g}(t)$. The forward pass consists of calculating timing as a function of arc-length through \eqref{time}, then interpolating the initial arbitrarily parameterized discrete trajectory to be evenly sampled in time. Once $\mathbf{C}(t)$ is acquired, $\mathbf{g}(t)$ can be calculated by differentiation of the trajectory with respect to time.

Interpolation presents an issue for back-propagation as the binning and resampling operation applied to the independent variables is not differentiable. The gradient is approximated then by assuming the operation is performed by a pair of constant copying matrices with no dependence on the optimization variables. The full backward pass through interpolation is given in greater detail in Appendix \ref{sec:appendix}.

\subsection{Time Domain Constraint Enforcement}
\begin{figure}[!t]
\centerline{\includegraphics[width=\columnwidth]{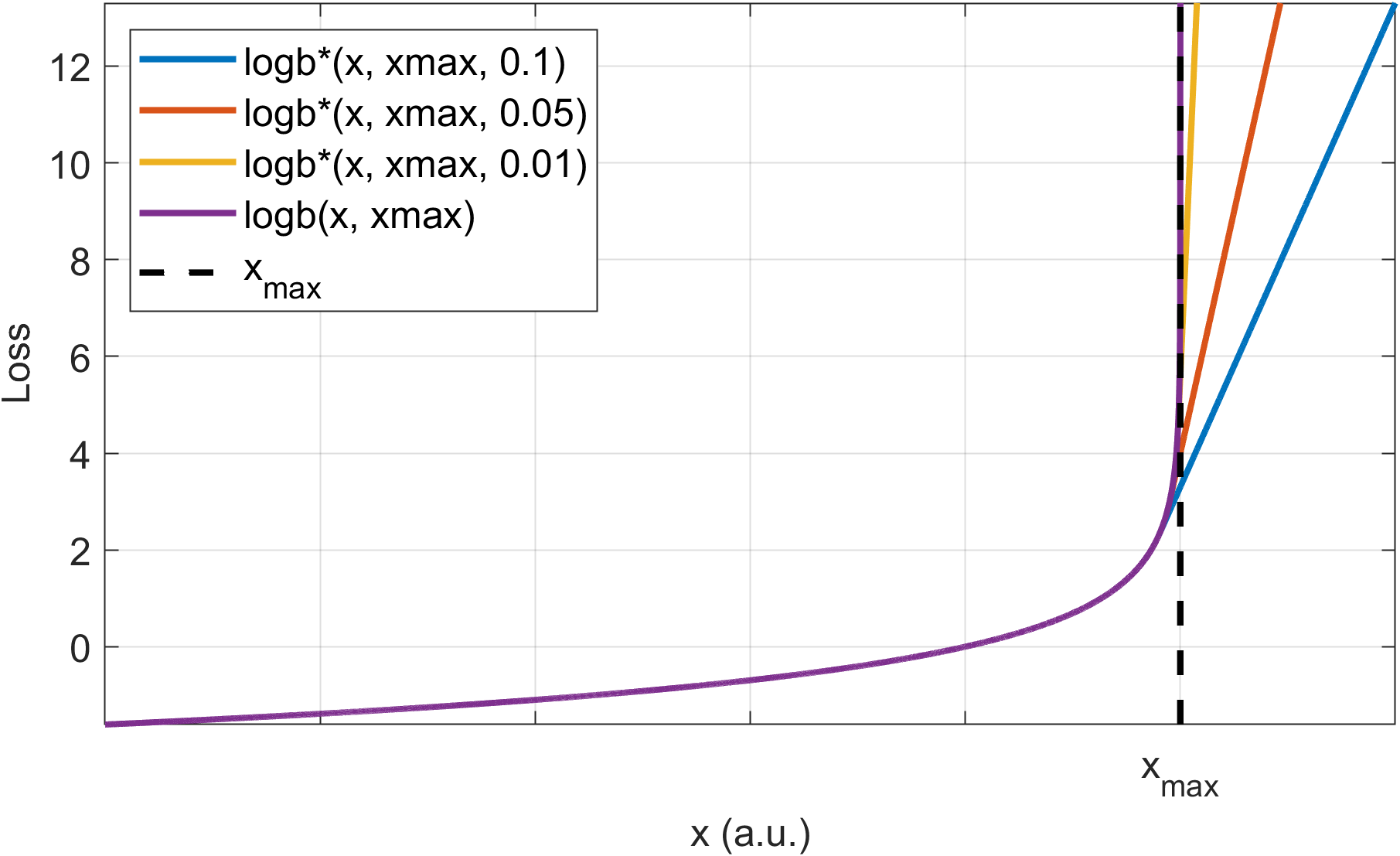}}
\caption{Comparison of log-barrier function (purple) to leaky log-barriers with varying relaxations. Smaller $\delta$ values result in steeper slopes beyond $x_{max}$.}
\label{fig:leaky_logb}
\end{figure}

Some waveform properties have a single limiting value which must be enforced as a constraint rather than general optimization. For example, waveforms are often limited by $P_{max}$ or $S_{max}$, or there may be some maximum duration $T_{max}$ we are willing to allow the waveform. Despite the non-convex nature of the problem presented here, we will apply a common approach to constraints used in convex problems. We introduce the log-barrier function \eqref{logb} which takes the loss asymptotically to infinity as input $x$ approaches its maximum allowed value $x_{max}$ \cite{boydcvx}
\begin{equation}
    \logb\left(x,\ x_{max}\right)=-\sum_{x} \ln{\left(x_{max}-x\right)}.
\label{logb}\end{equation}
If we compose the log-barrier \eqref{logb} with the total waveform duration $T=t(L)$ using \eqref{time}, we can enforce $T_{max}$ for the waveform design by minimizing $\logb(T,\ T_{max})$
\begin{equation}
    \logb\left(T,\ T_{max}\right)=-\ln{\left(T_{max} - \int_{0}^{L}{ds/v(s)} \right)}
\label{boundtime}\end{equation}
where we have dropped the summation given that $T$ is scalar. This bound time term \eqref{boundtime} is convex and increasing in argument $T$, while $T$ is convex in argument $v(s)$. We can therefore say that the bound time expression is convex in $v(s)$. However, for limiting time domain properties the use of a log-barrier presents an issue. Optimization in the arc-length domain while carried out in small steps can easily propagate to larger changes in $\mathbf{g}(t)$ properties which violate a log-barrier. This would result in infinite loss and halt the gradient descent process. To remedy this we introduce a “leaky” log-barrier function defined in \eqref{logb*}
\begin{equation}
\llogb(x,\ x_{max},\ \delta) = \begin{cases}
\logb\left(x,\ x_{max}\right) & x \le x_{\delta}\\
\sum_{x}\frac{x-x_{\delta}}{\delta}-\log{(\delta)} &\text{otherwise}.
\end{cases}
\label{logb*}\end{equation}
The leaky log-barrier function becomes linear beyond some point $x_{\delta}=x_{max}-\delta$ with continuous first derivative. The choice of $\delta$ determines the slope of the linear region, and therefore how hard the gradient descent is pushed back below the limit $x_{max}$ (Fig. \ref{fig:leaky_logb}). A similar relaxed approach has been applied in model predictive control with quadratic growth beyond the point $x_\delta$ \cite{llogb1, llogb2}.

\subsection{Time Optimal Gradient Descent}
We now have the tools to translate the minimum time problem \eqref{arcopt} into a form suitable for gradient descent. Enforcing the speed and speed-derivative constraints is necessary for designing feasible waveforms. The first constraint or “speed limit” can be enforced directly by introducing a placeholder variable $\xi(s)$, which defines $v(s)$ by \eqref{speedlim}
\begin{equation}
    \begin{aligned}
        v\left(s\right)\equiv\min{\left\{\gbar G_{max}\ ,\ \sqrt{\gbar S_{max}/\kappa\left(s\right)}\right\}}\cdot\sigma\left(\xi\left(s\right)\right).
    \end{aligned}
\label{speedlim}\end{equation}

Rather than optimizing $v(s)$ we optimize $\xi(s)$ and transform through \eqref{speedlim} where $\sigma$ is the sigmoid function. This trick is commonly used in machine learning to bound neural network layer outputs \cite{sigmoid}. This smoothly scales $v(s)$ to be between its lower bound 0 and the upper bound from \eqref{arcopt}, eliminating the speed limit constraint. However, the speed \textit{derivative} constraint forms a nonlinear differential inequality and cannot be directly enforced. The speed derivative constraint only relates to the $S_{max}$ limit and can be replaced with a leaky log-barrier loss term applied to slew-rate $\mathbf{S}(t)$ in the time domain. The result for the time-optimal gradient descent problem is given in \eqref{timeopt} where $\delta_S$ is the switching proximity to $S_{max}$ and $\lambda_1$ and $\lambda_2$ are relative weightings for the time minimization and slew-rate constraints respectively
\begin{equation}
\minimize_{\xi(s)} \quad \lambda_1\int_{0}^{L}{ds/v(s)} + \lambda_2\llogb(\lVert \mathbf{S}(t) \rVert_2,S_{max},\delta_S).
\label{timeopt}\end{equation}
The new optimization problem allows for the inclusion of additional design objectives as terms in the loss function. Minimizing \eqref{timeopt} with an iterative gradient descent approach requires an initial solution. In this work we initialize the OPTIKS method with the time optimal speed $v^*(s)$ for the given trajectory reduced by some factor $\alpha\in[0.8,1)$. The initial choice of $\xi(s)$ is found by inverting \eqref{speedlim} with $\alpha v^*(s)$.

The constraints on $v(s)$ and $v'(s)$ in \eqref{arcopt} ensure that $G_{max}$ and $S_{max}$ limits are maintained when $\mathbf{g}(t)$ is rotated \cite{miki_topt}. The resulting waveforms are therefore rotationally invariant, meaning that additional interleafs or oblique planes may be generated by simply rotating the existing solution $\mathbf{g}(t)$. This property is maintained in \eqref{timeopt} as the leaky log-barrier is applied to the slew-rate magnitude rather than a per channel basis, and the change of variables in \eqref{speedlim} directly enforces the invariant $G_{max}$ constraint.

\subsection{Peripheral Nerve Stimulation Control}
PNS is often more limiting in waveform design than system limits. We can impose a PNS limit $P_{max}$ on our waveforms by including an additional leaky log-barrier term in the optimization. In this work we use the standard PNS calculation set by the international electromechanical commission (IEC) as convolution of slew-rate with a nerve response function in \eqref{PNS} \cite{IEC60601}
\begin{equation}
    \begin{aligned}
        & h\left(t\right)\equiv\frac{\alpha c}{r\left(c+t\right)^2} \quad \forall t\geq0 \\
        & P\left(t\right)=100\lVert h(t) * \mathbf{S}(t) \rVert_2.
    \end{aligned}
\label{PNS}\end{equation}

The nerve response $h\left(t\right)$ is determined by system hardware, with rheobase $r$, chronaxie time $c$, and effective coil length $\alpha$ gradient coil specific constants provided by manufacturers. As the stimulation $P(t)$ is determined by a L2-norm across axes, the property is invariant under a rotation of the gradient waveform. Note that any PNS calculation which is differentiable with respect to $\mathbf{g}(t)$ could be used in place of \eqref{PNS} (e.g. the “SAFE” model \cite{safemodel}). Incorporating the PNS threshold limit into the optimization gives \eqref{PNSopt}
\begin{equation}
    \begin{aligned}
\minimize_{\xi(s)} \quad & \lambda_1\int_{0}^{L}{ds/v(s)} + \lambda_2\llogb(\lVert \mathbf{S}(t) \rVert_2,S_{max},\delta_S)\\ &+ \lambda_3\llogb(P(t), P_{max}, \delta_P).
    \end{aligned}
\label{PNSopt}\end{equation}

\subsection{Mechanical Resonance Control}
Vibration controls can only be implemented by passing through the time domain. The simplest design method for avoiding mechanical resonance is to identify the set of resonance frequency bands $\mathcal{M}$ and minimize power deposition within them as a term of the loss function \eqref{freqopt}
\begin{equation}
    \begin{aligned}
        \minimize_{\xi(s)} \quad & \lambda_1\int_{0}^{L}{ds/v(s)} + \lambda_2\llogb(\lVert \mathbf{S}(t) \rVert_2,S_{max},\delta_S)\\ &+ \lambda_3 \lVert \mathcal{F}\{\mathbf{g}(t)\}|_{f\epsilon\mathcal{M}} \rVert_F^2.
    \end{aligned}
\label{freqopt}\end{equation}

By including resonance bands from all 3 gradient axes in $\mathcal{M}$ and applying the Frobenius norm $\lVert\cdot\rVert_F$ we ensure that rotation of the OPTIKS waveforms will still avoid mechanical resonance. The waveforms can therefore be treated as rotationally invariant though rotation will produce slight changes in mechanical vibration. Another option is to directly limit the predicted oscillation either of field or physical displacement caused by the waveform. This requires the measurement of a vibration based impulse response function.

\begin{figure*}[!h]
\centering{\includegraphics[width=2\columnwidth]{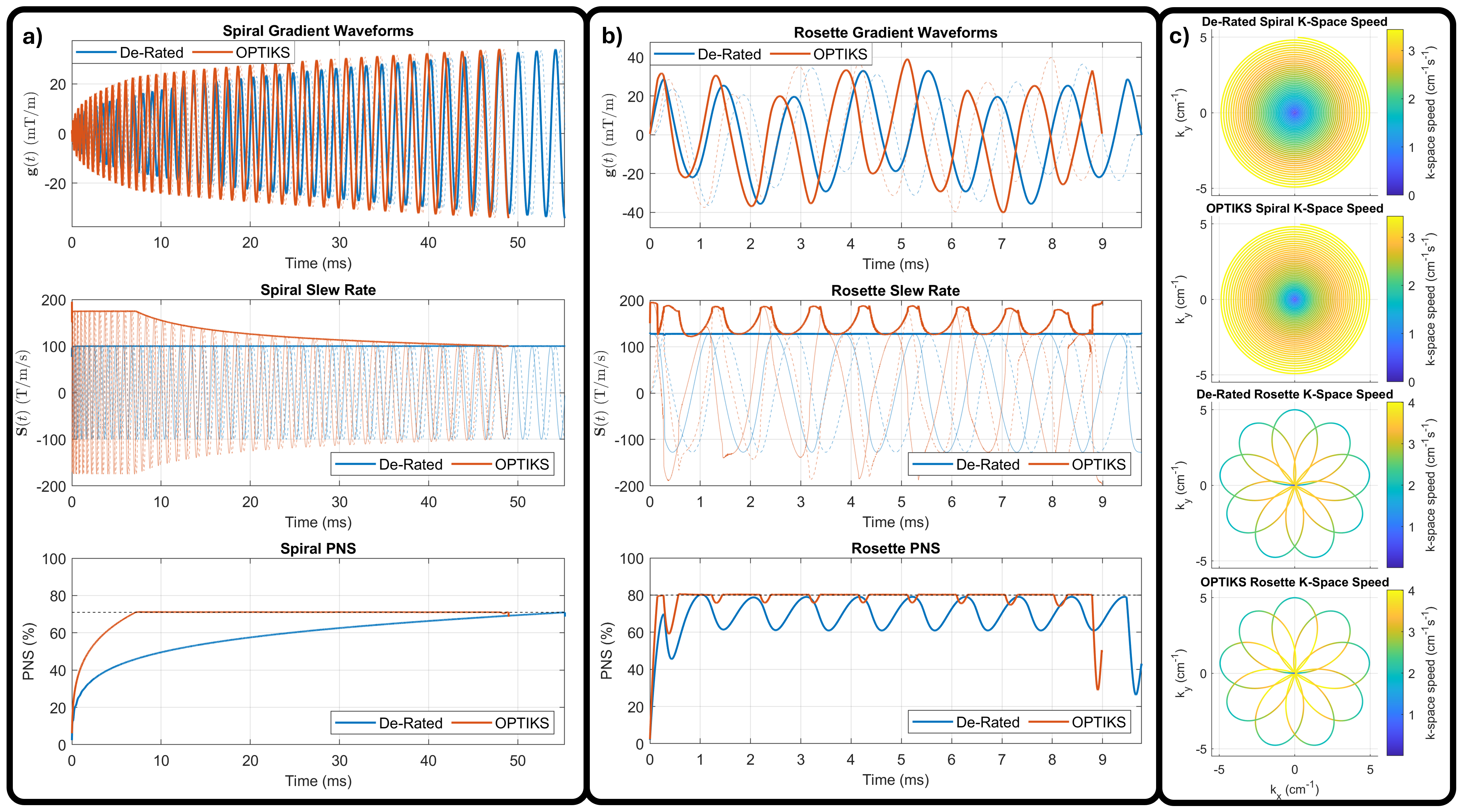}}
\caption{OPTIKS PNS-limited spiral and rosette designs for UHP. \textbf{a)} (TOP) Gradient waveforms for time-optimal (blue) and PNS-limited (orange) spiral (dashed Y channel). (MIDDLE) Slew-rate for the spiral waveforms, dark solid lines indicate magnitude across X and Y channels, faint solid and dashed lines indicate X and Y axes respectively. (BOTTOM) PNS response for spiral waveforms with dashed black line indicating $P_{max}$.  \textbf{b)} Gradient waveforms (TOP), slew-rate (MIDDLE), and PNS response (BOTTOM) for rosette designs. \textbf{c)} Resulting \textit{k}-space trajectories with color indicating speed along curve at each point. \rev{OPTIKS traverses k-space in approximately 10\% less time for the same PNS threshold in both cases.}}
\label{fig:PNS}
\end{figure*}

\subsection{Acoustic Noise Reduction}
Acoustic noise reduction is an extension of the mechanical resonance control problem. With a system-specific ATF the acoustic noise generated by a sequence can be predicted directly from the gradient waveforms. By including the predicted acoustic noise power as a term of the loss function we can design quieter waveforms \eqref{silent}
\begin{equation}
    \begin{aligned}
        \minimize_{\xi(s)} \quad & \lambda_1\int_{0}^{L}{ds/v(s)} + \lambda_2\llogb(\lVert \mathbf{S}(t) \rVert_2,S_{max},\delta_S)\\ &+ \lambda_3\lVert \mathbf{A}(f)\mathcal{F}\{\mathbf{g}(t)\}(f) \rVert_F^2.
    \end{aligned}
\label{silent}\end{equation}
Where $\mathbf{A}(f)$ is the ATF in the frequency domain. Minimizing acoustic power in the frequency domain is equivalent to minimizing power in the predicted acoustic waveform by Parseval's Theorem. Optimizing $\mathbf{g}(t)$ against its corresponding axis responses in this work (e.g. $A_x\mathcal{F}\{g_x\}, A_y\mathcal{F}\{g_y\}, A_z\mathcal{F}\{g_z\}$) locks the waveform into a specific orientation, allowing rotations of the design to maintain high acoustic output. If instead the tensor product across axes were optimized (e.g. $A_x\mathcal{F}\{g_x\}, A_y\mathcal{F}\{g_x\}, A_z\mathcal{F}\{g_x\}, A_x\mathcal{F}\{g_y\},...$) the design would reduce its response for any axis and could therefore be treated as rotationally invariant. Note that acoustic output would still change with orientation, but remain low.

\subsection{Resonance Characterization} \label{characterization}
To avoid vibrations at mechanical resonances, the resonance frequency bands were identified. Each gradient axis was probed by playing 120 ms duration sinusoids at every 10Hz from 50-2000 Hz and collecting field or back-EMF measurements for 20 ms following each waveform. Oscillatory error fields produced by lingering vibrations were monitored using a dynamic field camera (Skope, Zurich, Switzerland). The back-EMF produced by motion of the coils through the main magnetic field was also monitored using a digital scope (ADP3450, Diligent, Pullman, WA, USA). The ADP was connected to a test point on the current controlled gradient power amplifier which provides a scaled version of the bridge voltage reduced by a factor of 200x.

In both cases the measured oscillatory signals were represented by their RMS values at each input frequency to produce a spectrum of vibration intensity. Peaks in these spectra appear at the resonance bands and can be avoided in the gradient waveform design using \eqref{freqopt}.

\subsection{Acoustic Characterization}
ATFs were measured in-house for the GE 3T UHP and PREMIER systems to design OPTIKS waveforms with reduced acoustic noise. A pair of MR safe noise canceling headphones (OptoActive III, Optoacoustics Ltd., Mazor, Israel) with built-in calibrated microphones were placed 20 cm apart centered at isocenter on the patient table with the ear side facing up. The sound pressure level (SPL) was recorded during characterization waveforms played on each gradient axis. The audio output was connected to an Analog Discovery Pro for digitization. Sinusoids spaced every 10 Hz from 200-3130 Hz were each played and recorded for a duration of 160 ms. A reference waveform of 1000 Hz was recorded for each axis for a duration of 1 s.

The ATF for each system was fit using a least-squares solution of Fourier transformed input gradient waveforms $I_j(f)$ and output acoustic measurements $O_j(f)$ \eqref{ATF}

\begin{equation}
    A_i(f)=\frac{\sum\nolimits_{j}I^*_{i,j}(f) \cdot O_{i,j}(f)}{\sum\nolimits_{j}\lvert I_{i,j}(f)\rvert^2} \quad \forall i=x,y,z
\label{ATF}\end{equation}

and scaled to match the reference measurement at 1000 Hz.

\section{Results}
\label{sec:results}
In this section, we present examples of OPTIKS waveforms designed to limit PNS, minimize mechanical resonances, and reduce acoustic noise. We evaluate these designs based on theoretical PNS models, field camera measurements, back-EMF measurements, and acoustic measurements, and confirm that image quality is maintained with OPTIKS waveforms.

\begin{figure*}[!h]
\centering{\includegraphics[width=2\columnwidth]{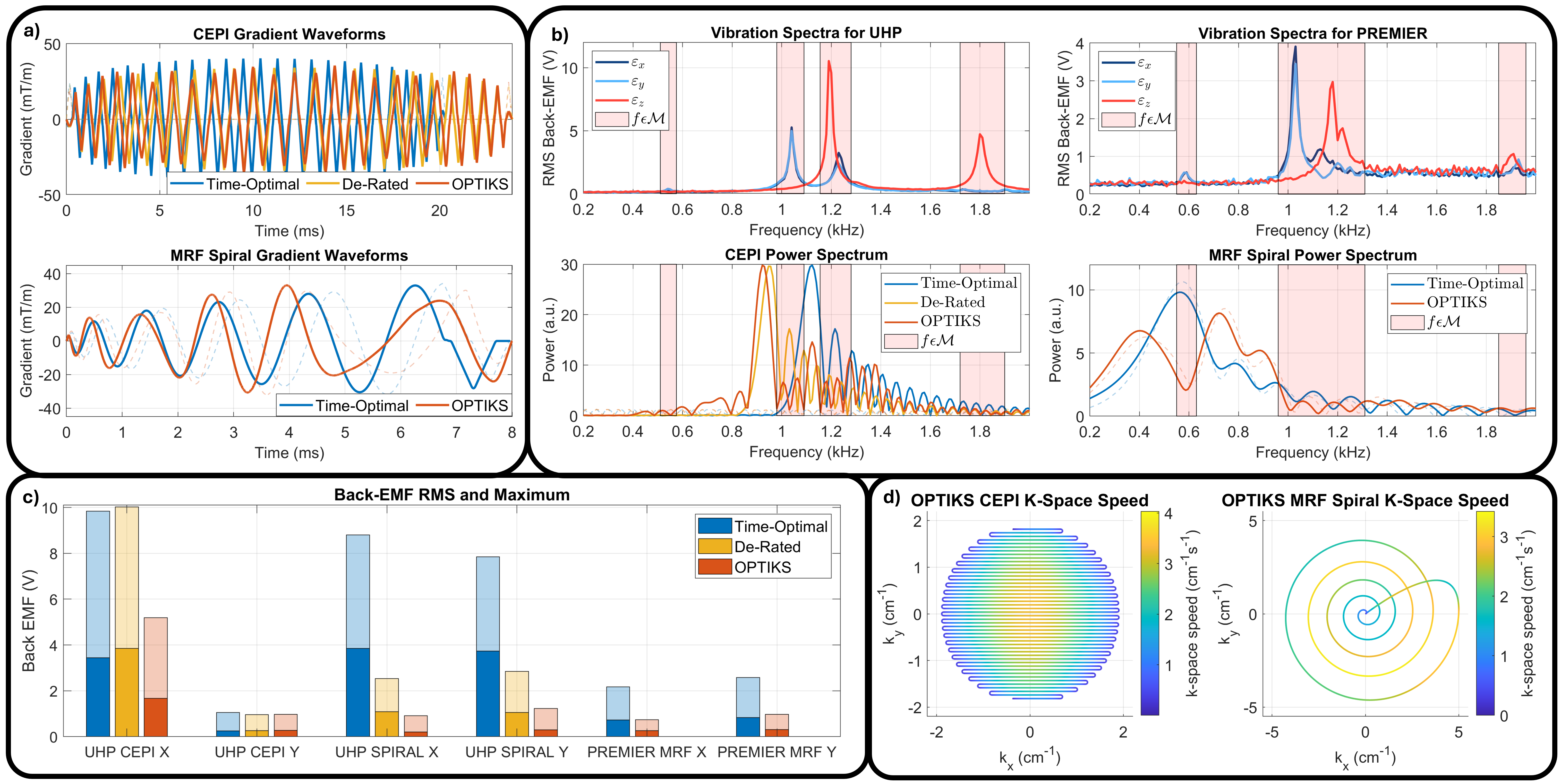}}
\caption{EMF mechanical resonance measurements and designs for UHP and PREMIER systems. \textbf{a)} Gradient waveforms for time-optimal (blue), de-rated (yellow), and OPTIKS mechanical resonance-optimized (orange) CEPI on UHP and MRF spiral on PREMIER. Solid and dashed lines representing X- and Y-axis data respectively.  \textbf{b)} Vibration spectra calculated from RMS back-EMF ($\varepsilon$) for UHP and PREMIER gradient systems, along with power spectra for CEPI and spiral waveforms. Mechanical resonances highlighted in red. \textbf{c)} RMS and maximum value of back-EMF measured from time-optimal, de-rated, and OPTIKS waveforms on UHP and PREMIER systems. Dark shorter bar gives RMS value, taller lighter bar gives maximum value. \textbf{d)} Resulting \textit{k}-space trajectories with color indicating speed along curve at each point for OPTIKS waveforms from b). \rev{In each design case OPTIKS significantly reduced power within mechanical resonance bands and the resulting vibration induced back-EMF.}}
\label{fig:EMF}
\end{figure*}

\subsection{Peripheral Nerve Stimulation Limited Waveforms}
PNS-limited waveforms were designed using OPTIKS for the GE 3T UHP system with the hardware $S_{max}$ of 195 T/m/s. These were compared to time-optimal waveforms designed using \eqref{arcopt} with globally de-rated $S_{max}$ to meet target $P_{max}$ values. Two in-house trajectory examples were considered; a 1 mm isotropic resolution 22 cm FOV variable density spiral previously used for diffusion weighted imaging \cite{diffspiral} and a 1 mm nine-petal rosette (Fig. \ref{fig:PNS}). 

The existing spiral waveform \cite{diffspiral} used a de-rated $S_{max}$ of 100 T/m/s reaching a $P_{max}$ of 72\% and the OPTIKS waveform was subsequently designed with the same target $P_{max}$. The de-rated rosette used a $S_{max}$ of 127.5 T/m/s to meet the "Normal Operating Mode" $P_{max}$ of 80\% set by IEC 60601-2-33 \cite{IEC60601}. The OPTIKS rosette was designed with the same $P_{max}$ of 80\%. The design functional consisted of time minimization, slew-rate limiting, and PNS limiting terms \eqref{PNSopt}. All waveforms used a $G_{max}$ of 100 mT/m.

For these specific trajectories and PNS limits the OPTIKS spiral requires a traversal time 6.31 ms shorter than its time-optimal counterpart, while the OPTIKS rosette is shorter by 0.79 ms. These equate to 11.4\% and 8.1\% reductions in traversal time respectively.

\subsection{Mechanical Resonances and Safe Waveforms} \label{mechresres}
Mechanical resonance excitation is particularly prominent in trajectories with periodic oscillations. Spiral and CEPI trajectories have gradient waveforms which sweep through a large range of frequencies, exciting multiple mechanical resonances. OPTIKS was employed to minimize mechanical resonance excitations from a CEPI waveform and single-shot spiral for the GE 3T UHP system, and a variable density magnetic resonance fingerprinting (MRF) spiral for the GE 3T PREMIER system \cite{MRF1, MRF2}. All waveforms were compared to their time-optimal counterparts designed with the same $S_{max}$. The single-shot spiral and CEPI waveforms were also compared to time-optimal waveforms with a de-rated $S_{max}$ chosen to achieve the same duration as the OPTIKS designs (Fig. \ref{fig:EMF}).

The CEPI trajectory was chosen to be 2.8 mm isotropic resolution with a 26 cm FOV and $R_y=2$. These trajectory parameters result in  gradient waveforms with power centered at the mechanical resonance bands in the time-optimal solution. The UHP spiral was designed with a 2 mm isotropic resolution and 24 cm FOV. All UHP waveforms were designed with $G_{max}=100$ mT/m. The time optimal and OPTIKS designs use $S_{max}=200$ T/m/s while the de-rated designs were reduced to $S_{max}=108$ T/m/s and $S_{max}=143$ T/m/s for the CEPI and spiral respectively. The OPTIKS designed CEPI waveforms reduced the RMS back-EMF by 70.8\% on the readout axis (X) and 55.2\% on the phase encode axis (Y), while the RMS oscillatory field was reduced by 56.3\% and 51.7\% respectively. The OPTIKS designed spiral waveforms reduced the RMS back-EMF by 94.8\% and the RMS oscillatory field by 91.1\%. These improvements came at the expense of speed with a 30\% and 15\% increase in duration respectively. Slew-derated CEPI waveforms reduced the RMS back-EMF by 36.7\% and 0.7\% on the readout and phase encode axes respectively, while the RMS oscillatory field was reduced by 46.2\% and 9.9\%. The slew-derated spiral waveforms reduced RMS back-EMF by 71.7\% and RMS oscillatory field by 69.8\%.

\begin{figure}[h]
\centering{\includegraphics[width=1\columnwidth]{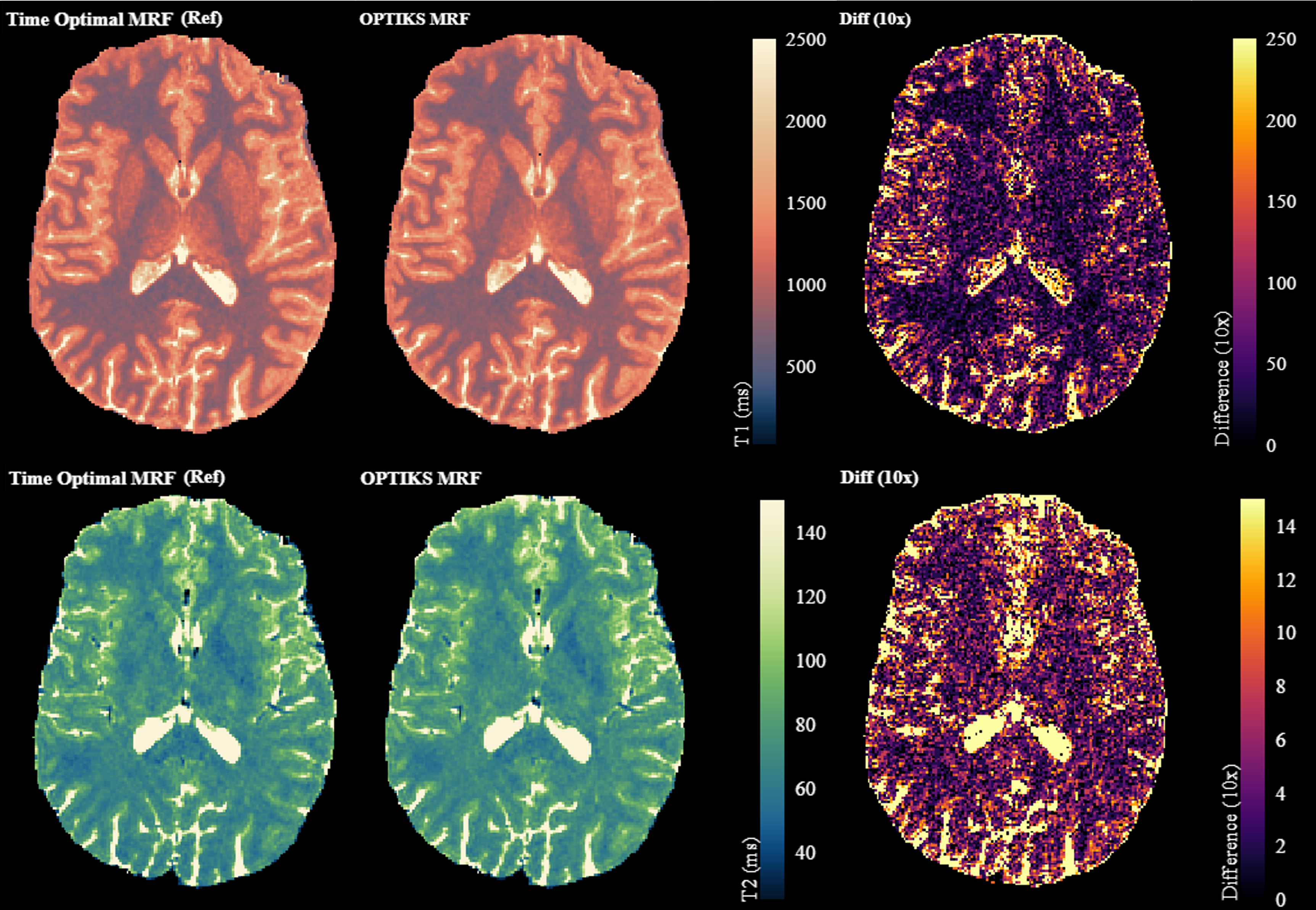}}
\caption{MRF results using time-optimal spiral and OPTIKS spiral. (TOP) Calculated T1 maps for each acquisition and their 10x difference map. (BOTTOM) Calculated T2 maps for each acquisition and their 10x difference map. \rev{OPTIKS maintained image quality and quantitative results within reported repeatability margins.}}
\label{fig:mrf}
\end{figure}

The spiral trajectory used for MRF in the PREMIER system was designed with a 1 mm isotropic resolution and a 22 cm FOV. A variable-density spiral design was employed, with the undersampling factor increasing linearly from R = 16 at the \textit{k}-space center to R = 32 at the end of the spiral interleaf, enabling faster acquisition. To ensure that the TR remained constant and thus preserve the consistency of the MRF dictionary, the OPTIKS spiral waveform was designed to match the duration of the original time-optimal waveform (8 ms). To offset the cost of mechanical resonance minimization, $S_{max}$ was increased from 100 T/m/s to 150 T/m/s and a PNS limit term \eqref{PNSopt} was included in the design with $P_{max}=80\%$. The time minimization term was replaced with a bound-time term \eqref{boundtime} of $T_{max}=8$ ms. The OPTIKS designed MRF spiral reduced the RMS back-EMF by 63.8\%. A field camera was not available for measurements on PREMIER. Reconstructed T1 and T2 maps from the MRF sequence are shown for the time-optimal and OPTIKS designed spiral acquisitions in Fig. \ref{fig:mrf}. Both images were acquired with informed consent from the subject.

\begin{figure}[!h]
\centering{\includegraphics[width=1\columnwidth]{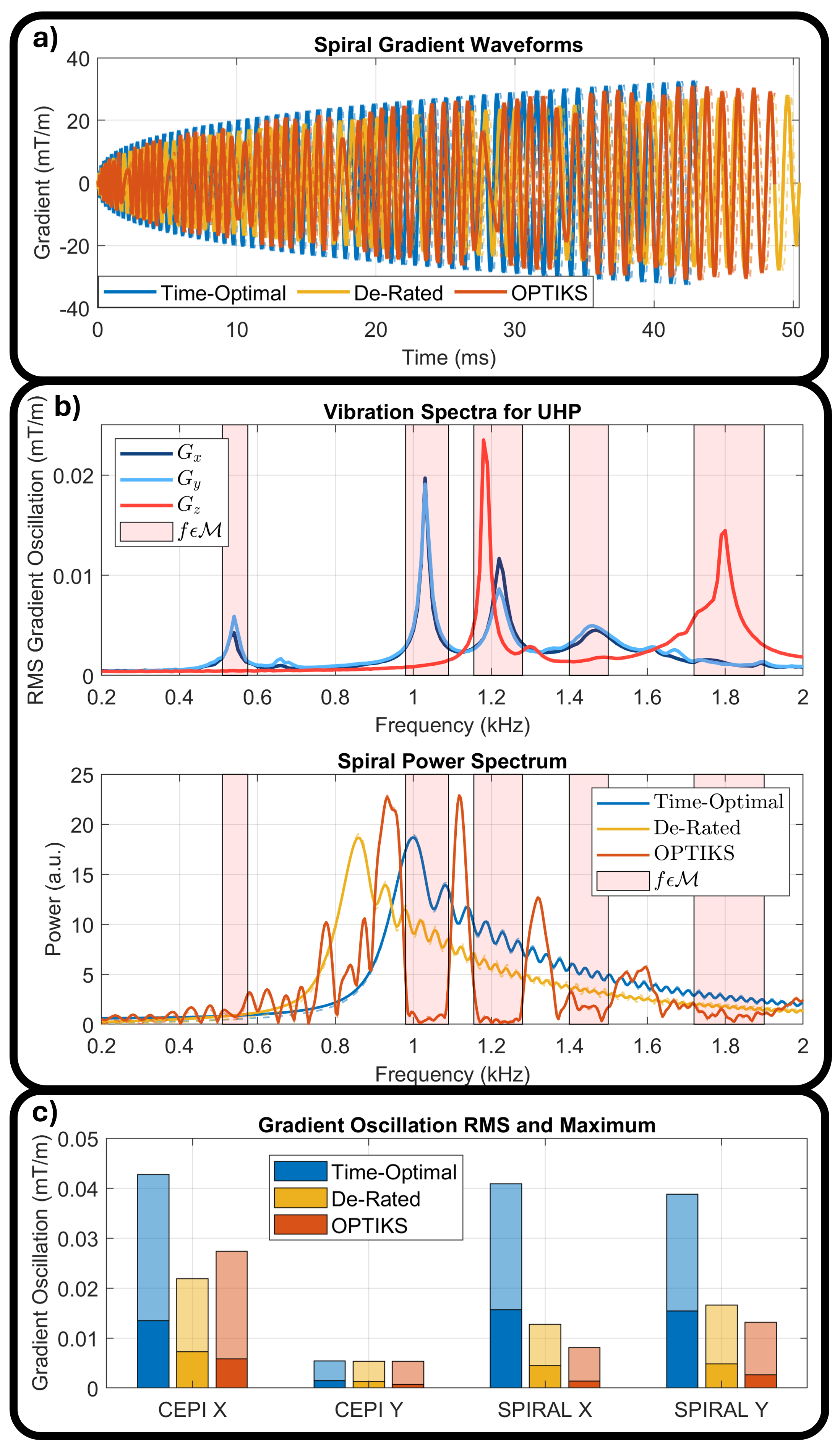}}
\caption{Field camera measurements and mechanical resonance spiral design. \textbf{a)} Gradient waveforms for time-optimal (blue), de-rated (yellow), and OPTIKS mechanical resonance-optimized (orange) single-shot spiral on UHP. \textbf{b)} (TOP) RMS gradient field oscillation spectra for the 3T GE UHP system on X (\rev{dark} blue), Y (\rev{light blue}), and Z (\rev{red}) axes. Mechanical resonance bands appear as peaks highlighted in red. (BOTTOM) Power spectra for time-optimal, de-rated, and OPTIKS single-shot spiral. \textbf{c)} RMS and maximum gradient field oscillations following time-optimal, de-rated, and OPTIKS waveforms played on the UHP system. Dark shorter bar gives RMS value, taller lighter bar gives maximum value. \rev{The OPTIKS spiral greatly reduced power within mechanical resonance bands, and vibration induced gradient oscillations were minimized for CEPI and spiral designs.}}
\label{fig:gosc}
\end{figure}

Mechanical resonance spectra obtained by back-EMF measurement for the UHP and PREMIER systems, and by field camera measurement for the UHP system are given in Fig. \ref{fig:EMF} and Fig. \ref{fig:gosc} respectively. Targeted mechanical resonance bands $\mathcal{M}$ are highlighted in red.

\begin{figure*}[!h]
\centering{\includegraphics[width=2\columnwidth]{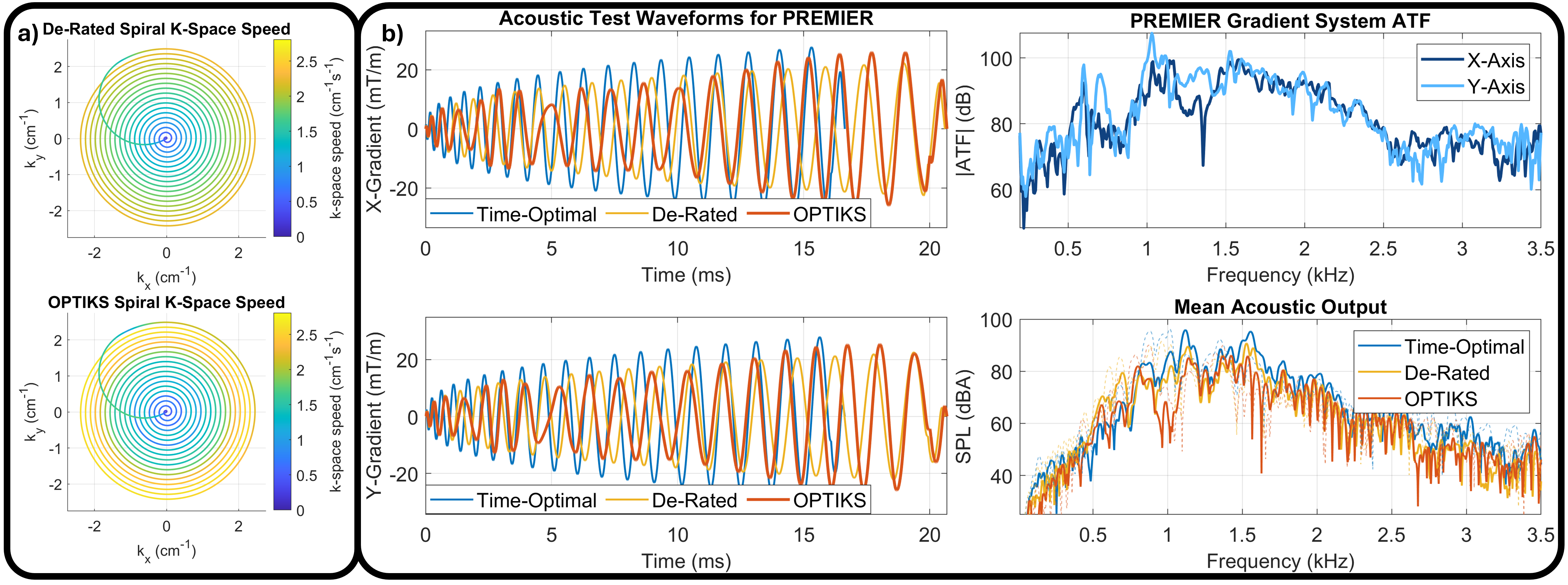}}
\caption{Acoustic noise reduction experiment. \textbf{a)} K-Space speed plots for de-rated and OPTIKS waveforms on the same colour axis. \textbf{b)} Test waveforms (LEFT) and measured outputs (BOTTOM-RIGHT). Time-optimal (blue), slew-de-rated (yellow), and acoustic optimized OPTIKS (orange) X and Y spiral waveforms designed for quiet imaging on PREMIER and measured output on PREMIER. Dark and lighter output lines indicate X and Y axes respectively.  Measured ATF for UHP and PREMIER gradient systems on X, and Y (TOP-RIGHT). \rev{OPTIKS reduced the mean acoustic output particularly at frequencies of high acoustic transference.}}
\label{fig:acoustic}
\end{figure*}

\subsection{Acoustic Characterization and Quieter Spirals}

The ATFs for the PREMIER and UHP systems are given in Fig. \ref{fig:acoustic}. Measurements are positionally dependenant and the ATF for each gradient axis was made up of the maximum measured values at each frequency for ATFs fit from two headphones. A $R=3$ rewound spiral with 2 mm isotropic resolution and 22 cm FOV was used for an acoustic minimization experiment on the PREMIER system. Three waveforms were designed for the spiral. A 16.628 ms time-optimal waveform and an OPTIKS waveform of 20.6 ms duration were designed with $S_{max}=135$ T/m/s. A waveform of equal duration 20.6 ms was designed to be time-optimal with slew-rate de-rated to $S_{max}=88$ T/m/s. The dBA weighted spectra of each recorded waveform is given in Fig \ref{fig:acoustic}. The decrease in peak output for each de-rated and OPTIKS waveform as compared to the time-optimal design is given in Table \ref{tab:acoustic}. OPTIKS gives a maximum noise decrease of 9.22 dB while the de-rated decreased noise by at most 5.37 dB with the same duration.

\begin{table}[h]
    \centering
    \begin{tabular}{c c c}
        \noalign{\hrule height 1.5pt}
        \multicolumn{3}{c}{Quiet Spiral Acoustic Noise Decrease on PREMIER System}\\
        \hline
        Coil Axis & De-Rated & OPTIKS\\
        \hline
        X-Axis Decrease (dBA) & 4.59$\pm$0.02 & 5.75$\pm$0.15\\
        Y-Axis Decrease (dBA) & 5.37$\pm$0.26 & 9.22$\pm$0.17\\
        \noalign{\hrule height 1.5pt}
    \end{tabular}
    \caption{Peak acoustic noise decreases from de-rated and OPTIKS spirals played on PREMIER as compared to time-optimal design.}
    \label{tab:acoustic}
\end{table}

\section{Discussion}
\label{sec:discussion}

We have shown that OPTIKS is a powerful tool for optimizing time-domain properties of fast arbitrary trajectory waveforms. Here we will discuss the trade-offs between OPTIKS and time-optimal waveforms, and interpret the solutions provided by this new design method.

\subsection{Peripheral Nerve Stimulation Performance}
\begin{figure}[h]
\centering{\includegraphics[width=1\columnwidth]{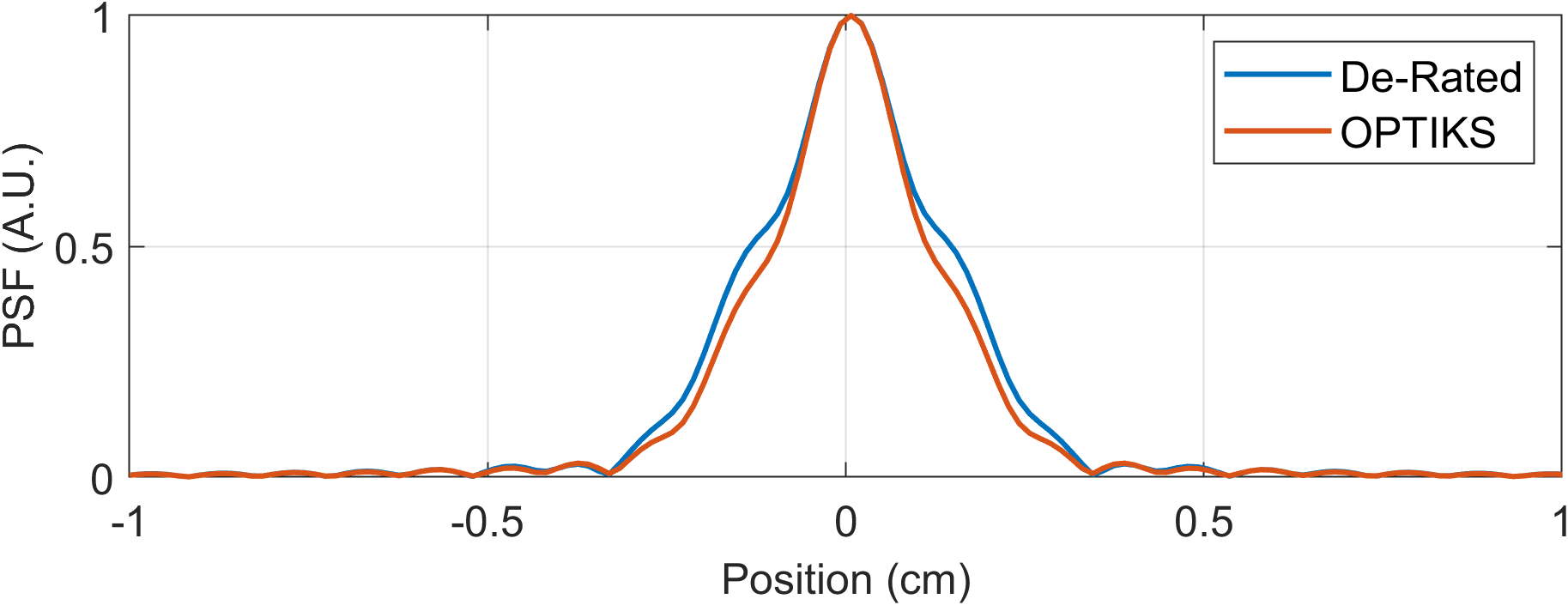}}
\caption{Point spread functions for slew de-rated and OPTIKS PNS-limited spiral designs. Simulated under T2\textsuperscript{*} decay and bulk off-resonance. Cropped to 1 cm radius to illustrate improved sharpness of OPTIKS readout. \rev{OPTIKS results in a narrower PSF.}}
\label{fig:PSF}
\end{figure}

OPTIKS was employed to design PNS-limited waveforms which minimize duration subject to $S_{max}$, $G_{max}$, and $P_{max}$ constraints. It was shown that the OPTIKS waveforms trade off slew-rate and PNS as limiting regimes in waveform design. The spiral waveform example produces a similar result to \cite{pns_spiral}, where the center of \textit{k}-space is slew-limited, and slew-rate is decreased when entering the PNS-limited regime. The speed boost for the spiral trajectory in the center of k-space would result in less T2\textsuperscript{*} decay and lower off resonance phase accumulation reducing subsequent blurring. An example point spread function (PSF) was simulated for each waveform given a typical gray matter T2\textsuperscript{*} of 50 ms and a bulk off-resonance of 10 Hz (Fig \ref{fig:PSF}). From the full-width-half-max of each PSF we can see that for this example the effective resolution of the time-optimal waveform is 2.82 mm while the OPTIKS waveform maintains a sharper image at 2.09 mm. The rosette trajectory also alternates between slew and PNS-limited regimes with each petal. In both cases the OPTIKS waveform exhibited a significant speed up over the naive slew de-rated waveform while meeting the same $P_{max}$. The OPTIKS design method improves on \cite{pns_spiral} by generalizing to arbitrary trajectories without adjustment as demonstrated with the rosette trajectory and rewound MRF spiral. As in \cite{pns_spiral} the design method optimizes a heuristic PNS \textit{model}. PNS is patient specific and may still be induced in some cases despite meeting model limits. Future work could investigate PNS caused by OPTIKS designs optimized with the IEC model \cite{IEC60601} or more sophisticated models such as SAFE \cite{safemodel}.

\subsection{Mechanical Resonance Performance}
OPTIKS waveforms were shown to drastically reduce the measured back-EMF and gradient field oscillations implying that mechanical resonance frequencies were avoided. The relative cost in speed was much lower for spiral designs than the CEPI trajectory. This is likely because the smooth spiral designs are easier to optimize without violating slew-rate constraints. Slew-rate in trajectories with regions of high curvature such as in CEPI would be very sensitive to increases in $v(s)$. The phase encode axis of the CEPI waveforms deposits very little power in the mechanical resonance bands without OPTIKS, but is still further reduced during the optimization.

The \textit{k}-space speed plots in Fig. \ref{fig:EMF} d) show how $v(s)$ was optimized along the trajectory to avoid resonances. In the case of the spiral the speed is decreased along some radii in a manner similar to the instantaneous frequency assumption of \cite{safespiral}. The improvement here relies on more subtle variation of the speed along the curve to avoid narrow resonance bands and transient effects, as well as the ability to include a rewinder in the trajectory. The CEPI design varies its speed for each line to avoid switching periods which fall directly into resonance bands or place harmonics in them. The MRF maps appear qualitatively similar but show an approximate 3\% and 5\% difference in T1 and T2 respectively for the grey and white matter, with higher differences localized to the cerebral spinal fluid. These differences are within the repeatability margins reported for the 2 minute MRF sequence used \cite{MRF3}.

\begin{figure}[h]
\centering{\includegraphics[width=1\columnwidth]{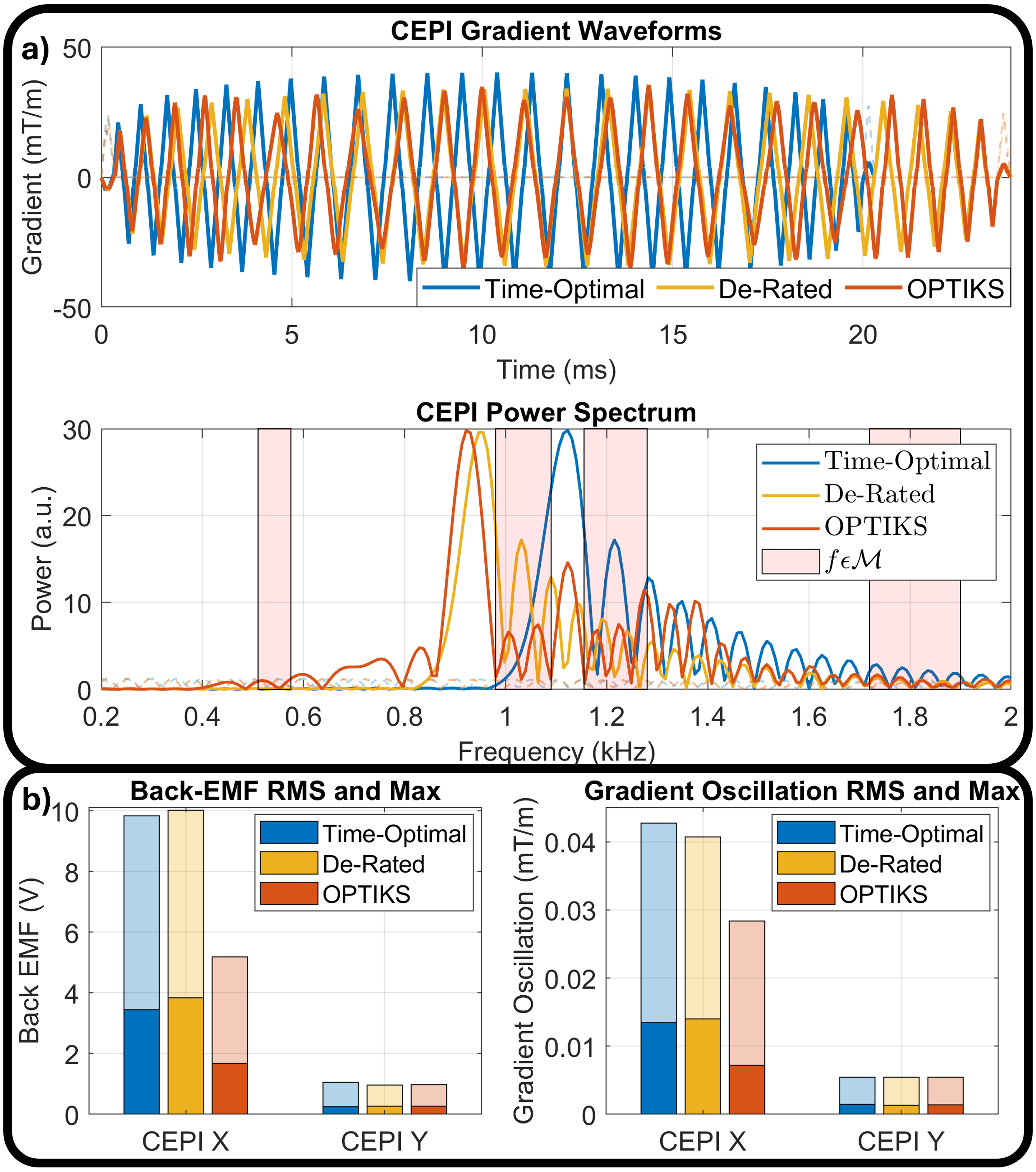}}
\caption{CEPI slew de-rated mechanical resonance avoidance fail case. \textbf{a)} Time-optimal (blue), de-rated (yellow), and OPTIKS (orange) CEPI gradient waveforms (TOP) and their spectra (BOTTOM). Mechanical resonance bands highlighted in red. \textbf{b)} RMS and maximum back-EMF and gradient oscillations measured for CEPI waveforms from a) on the UHP system. \rev{Naively de-rated designs increased mechanical resonance vibrations while OPTIKS reliably decreased vibrations.}}
\label{fig:counter}
\end{figure}

While the slew de-rated designs presented here reduced the back-EMF and field oscillations, decreasing slew-rate provides a very poor control for avoiding mechanical resonance. By globally reducing $S_{max}$ the waveform oscillations are elongated and the power spectrum is shifted towards lower frequencies without dramatic change to the spectra's shape. For a spiral waveform which exhibits a wide power spectrum the waveform would have to be significantly de-rated to reach the level of OPTIKS designs, with diminishing returns per unit time the waveform is elongated. While CEPI has a narrower spectrum a de-rated design suffers from the same limitations of the spiral. Furthermore, de-rating the time optimal waveform can shift the peak power deposition \textit{into} a mechanical resonance band if the user is not careful, while OPTIKS will seek to optimize the spectrum for any specified duration. As an example, we can design a new OPTIKS waveform for our CEPI trajectory from section \ref{mechresres} with a bound time constraint \eqref{boundtime} of $T_{max}=23.9$ ms. De-rating the time-optimal design to $S_{max}=133$ T/m/s for the same duration \textit{increases} the RMS back-EMF and gradient oscillation, as well as the maximum back-EMF (Fig. \ref{fig:counter}).

\subsection{Acoustic Noise Performance}
It has been shown that ATFs can have high positional dependence \cite{acoustic_spatial}. In this work two microphones were used, spaced equally apart from the scanner isocenter. At high frequencies these measurements diverged somewhat, likely due to formation of standing waves within the bore. To address this, the maximum of the two values was taken at each frequency for use in the acoustic noise minimization design \eqref{silent}. Additionally, the microphones used exhibit high directional dependence due to the noise-canceling headphone form-factor. The goal here is to provide a proof of concept for using OPTIKS to design quieter waveforms. The results of the acoustic spiral PREMIER test show a greater decrease in noise from the Y gradient axis than the X. This is likely due to more dramatic peaks and troughs in the Y axis ATF below 1 kHz, which the OPTIKS method takes advantage of for greater noise reduction. We showed that OPTIKS can out perform a naively de-rated waveform of equal duration by exploiting knowledge of system frequency response to band-limit itself and avoid frequencies of high transference. Further studies could expand upon this by use of an omnidirectional microphone and measurements over an array of positions.

\subsection{Optimality and Convergence}
The OPTIKS design functional is not convex and a given solution cannot be called globally optimal. The use of “leaky” log-barrier functions \eqref{logb*} to enforce constraints rewards reducing properties below the maximum allowed thresholds. \rev{A leaky rectified linear unit (ReLU) would lose the continuous first derivative of the “leaky” log-barrier resulting in large jumps in gradient direction and still reward reducing properties below thresholds.} A piecewise function with zero slope below the maximum threshold would avoid this \rev{over-reduction} issue, but would frequently violate limits during optimization due to the vanishing gradient. There is a trade off then between limit enforcement and operating efficiently close to the limit. Applying a small coefficient to these terms can push the slope towards zero below the threshold helping to reduce this issue.

Additionally, due to the problem's non-convexity and the relaxation of the barrier function it is possible for designs to slightly violate the slew-rate or PNS limits. When this occurs the limit can easily be reinforced by decreasing the hyperparameter $\delta$ in \eqref{logb*}.  Decreasing $\delta$ increases the slope beyond the threshold asymptotically towards infinity (Fig. \ref{fig:leaky_logb}) pushing the gradient strongly towards the allowed solution domain. Note that $\delta$ acts as a relaxation parameter, and that as $\delta$ approaches 0 we recover the true log-barrier function. Choosing $\delta$ to be arbitrarily small then can result in the same explosion of loss as the log-barrier function if the threshold is violated briefly during gradient descent. In this work slew limits were enforced with $\delta_S$ of $2\times10^{-4}$ and PNS limits were enforced with $\delta_P$ of $5\times10^{-5}$. Loss term weightings ($\lambda_i$) used in this work were by order of magnitude $\sim$10\textsuperscript{0} for bound time, $\sim$10\textsuperscript{4} for minimized time, $\sim$10\textsuperscript{2} for slew limits, $\sim$10\textsuperscript{1} for PNS limits, $\sim$10\textsuperscript{3} for mechanical resonance minimization, and $\sim$10\textsuperscript{5} for acoustic noise reduction.

\begin{figure}[h]
\centering{\includegraphics[width=1\columnwidth]{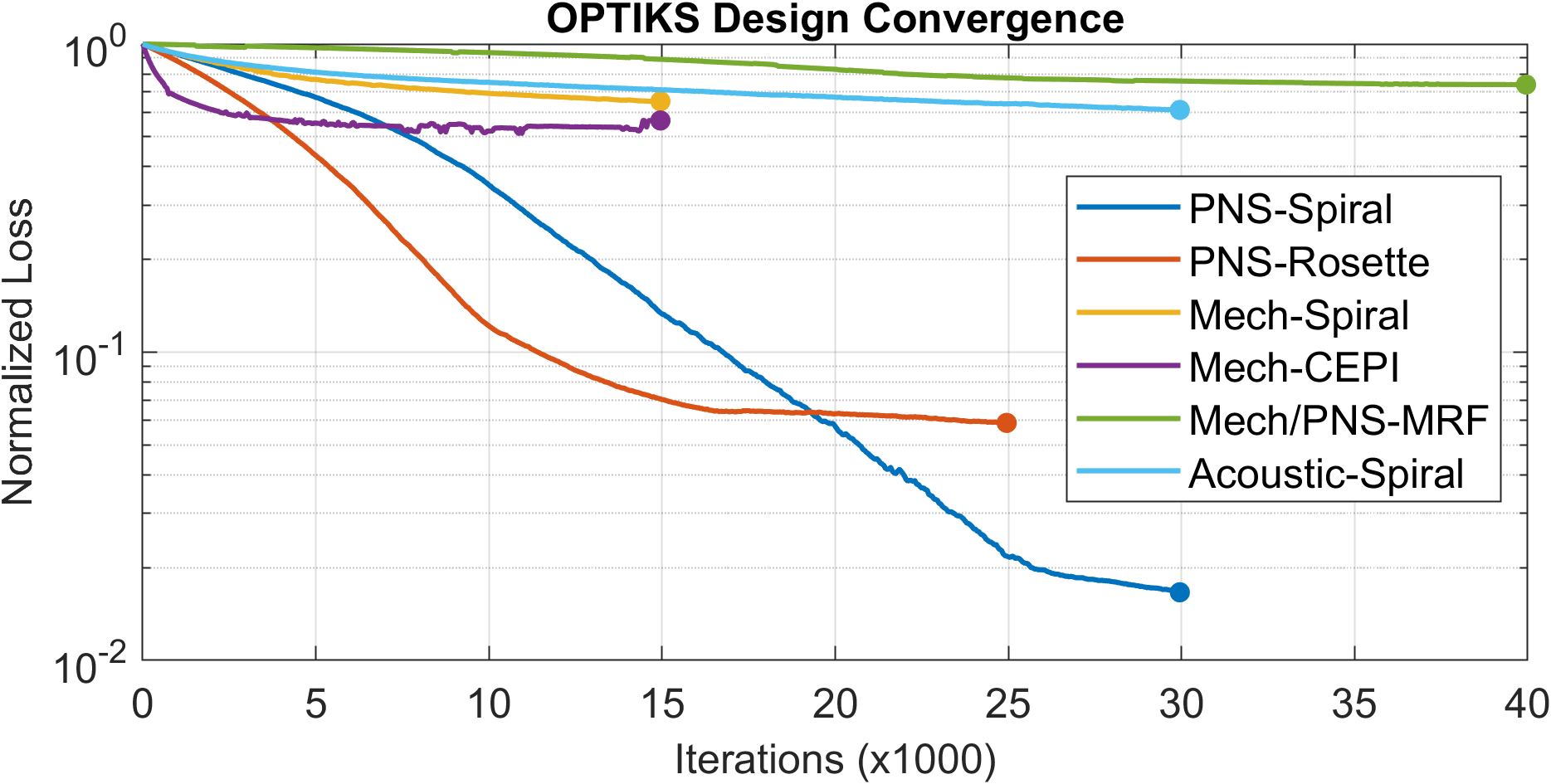}}
\caption{Convergence plots of normalized loss over SGD iteration for each OPTIKS design from section \ref{sec:results}. "Mech" refers to mechanical resonance minimized, and "acoustic" to acoustic noise reduced. Total wall clock computation times for OPTIKS as follows; PNS-Spiral 10 min 33 s, PNS-Rosette 8 min 29 s, Mech-Spiral 6 min 20 s, Mech-CEPI 5 min 56 s, Mech/PNS-MRF 15 min 15 s, Acoustic-Spiral 9 min 49 s.}
\label{fig:converge}
\end{figure}

Convergence plots for each of the 6 designs presented in this work are given in Fig. \ref{fig:converge}. The PNS limited spiral and rosette show the largest drop in loss as they were initialized with large PNS violations. The leaky log-barrier term enforcing PNS limits comes with a very steep gradient producing a more dramatic convergence plot than other design objectives. No convergence criteria were specified, and optimization was terminated by the user on a per case basis when a design was deemed acceptable. For PNS limited designs, a good convergence criterion would be to terminate optimization once the design is always either slew or PNS limited. Mechanical resonance and acoustic designs could be terminated once power falls below some minimum level in the resonance bands and predicted acoustic output. Future work could include determining acceptable power levels. The optimization method could also be improved using more sophisticated methods such as quasi-Newton methods like limited-memory BFGS.

\section{Conclusion}
\label{sec:conclusion}

A novel design method for optimization of time-domain gradient properties along arbitrary trajectories was developed, and made available as an open source Python package (https://github.com/mamccready/optiks). The OPTIKS method is highly flexible to customized loss functions, opening up trajectory constrained time-efficient gradient design to a host of new design goals. OPTIKS waveforms were shown to reduce readout time in PNS-limited scans, design quieter spiral waveforms, and avoid gradient mechanical resonances. These waveforms have the potential to increase patient comfort and reduce system strain while maintaining image quality in efficient time.

\appendices

\section{Interpolation Backward Pass}
\label{sec:appendix}

Linear interpolation of a pair of vectors $\mathbf{x},\mathbf{y}\epsilon\mathbb{R}^{m\times1}$ to a pair $\mathbf{p},\mathbf{q}\epsilon\mathbb{R}^{n\times1}$ consists of binning the independent values $\mathbf{p}$ into $\mathbf{x}$ and averaging the surrounding dependent values from $\mathbf{y}$ based on their distance to the new point. An example is given in \eqref{interpolation} where $\left(x_{j-1},\ y_{j-1}\right) and \left(x_j,\ y_j\right)$ surround the new point $\left(p_i,\ q_i\right)$

\begin{equation}
    q_i=y_{j-1}\left(\frac{x_j-p_i}{x_j-x_{j-1}}\right) +y_j\left(\frac{p_i-x_{j-1}}{x_j-x_{j-1}}\right).
\label{interpolation}\end{equation}

We define two permutation matrices $P_1,P_2\epsilon\mathbb{R}^{n\times m}$ which copy select values in $\mathbf{x}$ and $\mathbf{y}$ such that the surrounding values align with the new points \eqref{binning} and define a third matrix $P_3\equiv P_2-P_1$

\begin{equation}
    \left(P_1\mathbf{x}\right)_i\le p_i \le\left(P_2\mathbf{x}\right)_i \quad {\forall}i=1,...,n.
\label{binning}\end{equation}

The forward pass of interpolation can then be written using the original and interpolant vectors as in \eqref{vecinterpolation}

\begin{equation}
    \begin{aligned}
        \mathbf{q}=&\diag\left(P_1\mathbf{y}\right)\left(\frac{P_2\mathbf{x}-\mathbf{p}}{P_3\mathbf{x}}\right)\\
        &+\diag\left(P_2\mathbf{y}\right)\left(\frac{\mathbf{p}-P_1\mathbf{x}}{P_3\mathbf{x}}\right).
    \end{aligned}
\label{vecinterpolation}\end{equation}

The construction of the sorting matrices is not differentiable and cannot be included in the backpropagation algorithm. However, if we treat them as constant at each gradient descent step, we can approximate the backward pass as in \eqref{difftrans}

\begin{equation}
    \begin{aligned}
        &\mathbf{\nabla}_\mathbf{x}\mathbf{q}\approx \diag\left(\frac{P_1\mathbf{y}}{P_3\mathbf{x}}\right)\left[{P_2}^T-\diag\left(\frac{P_2\mathbf{x}-\mathbf{p}}{P_3\mathbf{x}}\right){P_3}^T\right] \\
        & \quad\quad + \diag\left(\frac{P_2\mathbf{y}}{P_3\mathbf{x}}\right)\left[\diag\left(\frac{\mathbf{p}-P_1\mathbf{x}}{P_3\mathbf{x}}\right){P_3}^T-{P_1}^T\right] \\
        &\mathbf{\nabla}_\mathbf{p}\mathbf{q}\approx \diag\left(\frac{P_3\mathbf{y}}{P_3\mathbf{x}}\right).
    \end{aligned}
\label{difftrans}\end{equation}

\bibliographystyle{ieeetr}
\bibliography{refs}

\end{document}